\documentclass[12pt,preprint]{aastex}

 \usepackage{epstopdf}
 \usepackage{caption}

 \slugcomment{Accepted Version 1}
 
 \newcommand\pv{\mbox{$p_{V}$}}
 \newcommand\irfactor{\mbox{$p_{IR}/p_{V}$}}

\begin{document}

 \DeclareGraphicsExtensions{.pdf,.gif,.jpg}

 \title{Thermal Model Calibration for Minor Planets Observed with WISE/NEOWISE: Comparison with IRAS}
 \author{A. Mainzer\altaffilmark{1}, T. Grav\altaffilmark{2}, J. Masiero\altaffilmark{1}, J. Bauer\altaffilmark{1}$^{,}$\altaffilmark{3}, E. Wright\altaffilmark{4}, R. M. Cutri\altaffilmark{3}, R. Walker\altaffilmark{5}, R. S. McMillan\altaffilmark{6}}
 \altaffiltext{1}{Jet Propulsion Laboratory, California Institute of Technology, Pasadena, CA 91109 USA}
 \altaffiltext{2}{Johns Hopkins University, Baltimore, MD}
\altaffiltext{3}{Infrared Processing and Analysis Center, California Institute of Technology, Pasadena, CA 91125, USA}
\altaffiltext{4}{UCLA Astronomy, PO Box 91547, Los Angeles, CA 90095-1547 USA}
\altaffiltext{5}{Monterey Institute for Research in Astronomy, Monterey, CA USA}
\altaffiltext{6}{Lunar and Planetary Laboratory, University of Arizona, 1629 East University Blvd., Kuiper Space Science Bldg. \#92, Tucson, AZ 85721-0092, USA}

 \email{amainzer@jpl.nasa.gov}

 \begin{abstract}
With thermal infrared observations detected by the NEOWISE project, we have measured diameters for 1742 minor planets that were also observed by the \emph{Infrared Astronomical Satellite} (IRAS).  We have compared the diameters and albedo derived by applying a spherical thermal model to the objects detected by NEOWISE and find that they are in generally good agreement with the IRAS values. We have shown that diameters computed from NEOWISE data are often less systematically biased than those found with IRAS.  This demonstrates that the NEOWISE dataset can provide accurate physical parameters for the $>$157,000 minor planets that were detected by NEOWISE.   
 
 \end{abstract}
 
 \section{Introduction}
Understanding the formation and evolution of the minor planets in our Solar System requires good knowledge of the physical parameters that describe each population of asteroids and comets.  Models of asteroid migration, for example, are critically dependent on measurements of number counts, sizes and albedos.  Visible light surveys have supplied most of the discoveries of asteroids and comets to date; however, these surveys tend to be biased against low albedo objects, and so they are underrepresented in number counts.  Furthermore, diameters derived from visible light observations are highly uncertain due to their linear dependence on albedo, which can vary by an order of magnitude for many types of asteroids.  Radar observations, stellar occultations and in situ spacecraft imaging provide powerful means of obtaining precise diameter measurements, but these observations are limited to a small subset of the $\sim$500,000 minor planets known to exist today.  While diameter and albedo can be measured more precisely than with visible light by applying radiometric models to objects for which thermal infrared data have been obtained, the size and albedo distributions obtained in this way will still be biased if the underlying source population was discovered by visible light surveys.  In order to mitigate against these biases, it is necessary to undertake a survey that is capable of independent detection and discovery of asteroids at thermal infrared wavelengths.  Yet the physical parameters such as size and albedo that are derived from such a new survey must be checked against other well-characterized datasets to ensure their reliability. 

The \emph{Wide-field Infrared Survey Explorer} (WISE) is a NASA Medium-class Explorer mission designed to survey the entire sky in four infrared wavelengths, 3.4, 4.6, 12 and 22 $\mu$m (denoted $W1$, $W2$, $W3$, and $W4$ respectively) \citep{Wright, Liu, Mainzer}.  The final mission data products are a multi-epoch image atlas and source catalogs that will serve as an important legacy for future research.   The portion of the pipeline dedicated to finding minor planets (called NEOWISE) has yielded observations of over 157,000 minor planets, including Near-Earth Objects (NEOs), Main Belt Asteroids (MBAs), comets, Hildas, Trojans, Centaurs, and scattered disk objects \citep{Mainzer11a}.  This represents an improvement of nearly two orders of magnitude more objects observed than WISE's predecessor mission, the \emph{Infrared Astronomical Satellite} \citep[IRAS;][]{Tedesco, Tedesco92, Matson}.  The WISE survey began on 14 January, 2010, and the mission exhausted its primary tank cryogen on 5 August, 2010.  Exhaustion of the secondary tank and the start of the NEOWISE Post-Cryogenic Mission occurred on 1 October, 2010, and the survey ended on 31 January, 2011.

In \citet{Mainzer11b}, we demonstrated that thermal models created for minor planets using the WISE/NEOWISE dataset produce results that are in good agreement with diameters independently measured by radar, stellar occultations, and in situ spacecraft imaging.  By taking the previously measured diameters from methods derived independently of infrared thermal model assumptions, we were able to verify the accuracy of the color corrections given in \citet{Wright} for low temperature objects such as asteroids.  We used the previously measured diameters combined with the \citet{Wright} color corrections in conjunction with a faceted spherical thermal model based on the Near-Earth Asteroid Thermal Model \citep{Harris} to predict the WISE magnitudes for $\sim$50 asteroids; these were compared with the observed WISE magnitudes.  We found that there were no systematic offsets between predicted and observed magnitudes for these objects, indicating that the color corrections given in \citet{Wright} adequately describe the system response.  For objects with WISE measurements in two or more bands with good signal-to-noise (for which the beaming parameter $\eta$ can be fit), we found that diameters can be determined to within $\pm$10\%, and visible albedo \pv to within $\pm$20\%.  

In this paper, we compare diameters and \pv\ obtained with WISE/NEOWISE observations for 1742 asteroids to those determined by IRAS and find that they are in generally good agreement with some slight systematic differences.   This work, combined with \citet{Mainzer11b}, demonstrates that the NEOWISE observations of minor planets will produce good physical parameters that will enable a wide range of scientific investigations. 

 \section{Observations}
We have assembled a list of objects that were observed by IRAS that NEOWISE detected during the fully cryogenic portion of its mission.   Of the $\sim$2200 asteroids observed by IRAS \citep{Tedesco, Tedesco92, Matson}, we identified NEOWISE detections for 1742 objects.  The observations of these objects were retrieved by querying the Minor Planet Center's (MPC) observation files to look for all instances of individual WISE detections of the desired objects that were reported using the WISE Moving Object Processing System \citep[WMOPS;][]{Mainzer11a}.  The resulting set of position/time pairs were used as the basis of a query of WISE source detections in individual exposures (also known as ``Level 1b" images) using the Infrared Science Archive (IRSA).  In order to ensure that only observations of the desired moving object were returned from the query, the search radius was restricted to 0.3 arcsec from the position listed in the MPC observation file.  Additionally, since WISE collected a single exposure every 11 seconds and observes each part of the sky an average of 10 times, the modified Julian date was required to be within 2 seconds of the time specified by the MPC.  The following flag values were allowed: cc\_flags = 0, P or p  and ph\_qual = A, B, or C.  Objects brighter than $W3=4$ and $W4=0$ magnitudes were assumed to have flux errors equivalent to 0.2 magnitudes due to changes to the shape of the point spread function as the objects became saturated, and a linear correction was applied to the $W3$ magnitudes in this brightness regime (the WISE Explanatory Supplement contains a more detailed explanation).  Per the Explanatory Supplement, objects brighter than $W3=-2$ and $W4=-6$ were not used.  Each object had to be observed a minimum of three times in at least one WISE band, and it had to be detected at least 40\% of the time when compared to the band with the maximum number of detections (usually, though not always, $W3$). The WMOPS system is designed to reject inertially fixed objects such as stars and galaxies in bands $W3$ and $W4$. Nonetheless, the individual images at all wavelengths were compared with WISE atlas coadd and daily coadd source lists to ensure that inertially-fixed sources such as stars and galaxies were not coincident with the moving object detections.  This check is particularly important in bands $W1$ and $W2$ where the density of background objects (and hence the probability of a blended source) is higher than at longer wavelengths.  Any remaining blended sources in bands $W1$ and $W2$ were removed.  Some objects were observed at multiple epochs, and observations separated by more than three days were modeled separately.  

\section{Thermal Model and Reflected Sunlight Fits}
 In the Standard Thermal Model (STM) of \citet{Lebofsky_Spencer}, the temperature of an asteroid is assumed to be maximum at the subsolar point and zero on the point opposite of this; this is the case of an object with zero thermal inertia.  In contrast, in the Fast Rotating Model (FRM) \citep{Lebofsky78,Veeder89,Lebofsky_Spencer}, the asteroid is assumed to be rotating much faster than its cooling time, resulting in a constant surface temperature across all longitudes.  The so-called ``beaming parameter" ($\eta$) was introduced by \citet{Lebofsky} in the STM to account for the enhancement of thermal radiation observed at small phase angles.  The near-Earth asteroid thermal model (NEATM) of \citet{Harris} also uses the beaming parameter $\eta$ to account for cases intermediate between the STM and FRM models.  In the STM, $\eta$ is set to 0.756 to match the occultation diameters of (1) Ceres and (2) Pallas, while in the FRM, $\eta$ is equal to $\pi$.  With NEATM, $\eta$ is a free parameter that can be fit when two or more infrared bands are available.

We modeled each object as a set of triangular facets covering a spherical surface with diameter equal to the ground-truth measurement \citep[c.f.][]{Kaasalainen}.  The temperature for each facet was computed, and the \citet{Wright} color corrections were applied to each facet.   The emitted thermal flux for each facet was calculated using NEATM along with the bandcenters and zero points given in \citet{Wright}; nightside facets were assumed to contribute no flux.  The emissivity, $\epsilon$, was assumed to be 0.9 for all wavelengths \citep[c.f.][]{Harris09}.  The objects' absolute magnitudes ($H$) were taken from \citet{Warner} when available; otherwise, the values were taken from the MPC's orbital element files.  Unless a direct measurement of $G$ was available from \citet{Warner}, we assumed a $G$ value of 0.15.  In general, minor planets detected by NEOWISE in bands $W1$ and $W2$ contain a mix of reflected sunlight and thermal emission.  Thus, it was necessary to incorporate an estimate of reflected sunlight into the thermal model in order to use data from bands $W1$ and $W2$.  In order to compute the fraction of reflected sunlight in bands $W1$ and $W2$, it was also necessary to compute the ratio of the albedo at these wavelengths compared to the visible abledo (\irfactor); the solar spectrum was approximated as a 5778 K blackbody.  As described in \citet{Mainzer11b}, we assumed that $p_{IR} = p_{3.4\mu m} = p_{4.6\mu m}$.  The flux from reflected sunlight was computed for each WISE band using the IAU phase curve correction of \citet{Bowell}. Thermal models were computed for each object by grouping together observations no more than a three day gap between them. 

Error bars on the model magnitudes and subsolar temperatures were determined for each object by running 25 Monte Carlo (MC) trials that varied the objects' $H$ values by 0.3 magnitudes and the WISE magnitudes by their error bars using Gaussian probability distributions.  The minimum magnitude error for all WISE measurements fainter than $W3=4$ and $W4=3$ magnitudes was 0.03 magnitudes, per the in-band repeatability measured in \citet{Wright}. For objects brighter than $W3=4$ and $W4=3$, the error bars were increased to 0.2 magnitudes, as these magnitudes represent the onset of saturation (see the WISE Explanatory Supplement). For those objects for which $\eta$ and \irfactor\ could not be fitted, $\eta$ was set to 1.0 and allowed to vary in the MC trials by 0.25, and \irfactor\ was set to 1.4 and allowed to vary by 0.5.  These default values were used because they are the mean and standard deviation of $\eta$ and \irfactor\ for the objects for which these parameters could be fit. The error bar for each object's model magnitude was equal to the standard deviation of all the MC trial values.  

Although many asteroids are known to be non-spherical, the WISE observations generally consisted of $\sim$10-12 observations per object uniformly distributed over $\sim$36 hours \citep{Wright, Mainzer11a}, so on average, a wide range of rotational phases was sampled.  Although the variation in effective spherical diameter resulting from rotational effects tends to be averaged out, caution must be exercised when interpreting effective diameter results using spherical models for objects that are known to have large-amplitude lightcurve variations.  In our sample, 179 objects have NEOWISE observations at multiple epochs (meaning that the groups of observations were separated by more than 10 days).  Of these, all but 24 had diameters that agreed to within 10\%, and most of the remaining objects had $W3$ peak-to-peak amplitudes $>$0.3 mag, indicating that they are likely to be non-spherical.  Non-spherical objects observed at different viewing geometries can lead to different diameters when using a spherical model. However, since 87\% of the objects with multi-epoch observations have diameters that agree to within 10\%, we conclude that multiple observational epochs do not contribute significantly to the differences observed between IRAS and NEOWISE diameters.  

\section{Results}
\begin{figure}
\figurenum{1}
\includegraphics[width=6in]{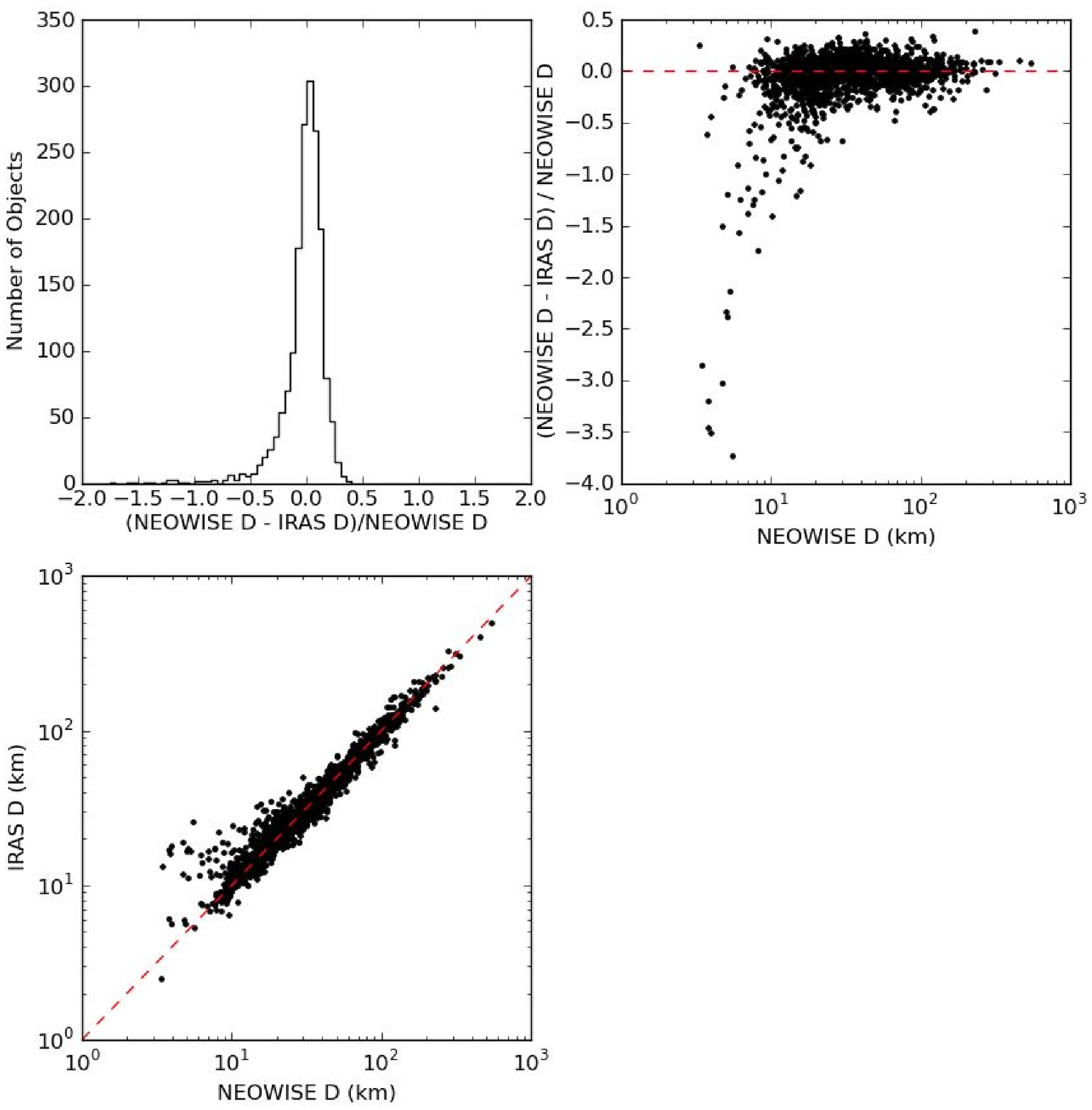}
\caption{\label{fig:iras_diameter_diff}Comparison of the difference in diameters given by \citet{TedescoPDS} and the diameters found by applying a spherical NEATM model to NEOWISE measurements for 1742 asteroids.  It can be seen that the values from \citet{TedescoPDS} tend to be slightly smaller than NEOWISE-derived diameters. The NEOWISE diameters diverge widely from IRAS for the smallest objects; however, these objects are observed by NEOWISE with high signal-to-noise ratio.   
}
\end{figure}

\begin{figure}
\figurenum{2}
\includegraphics[width=6in]{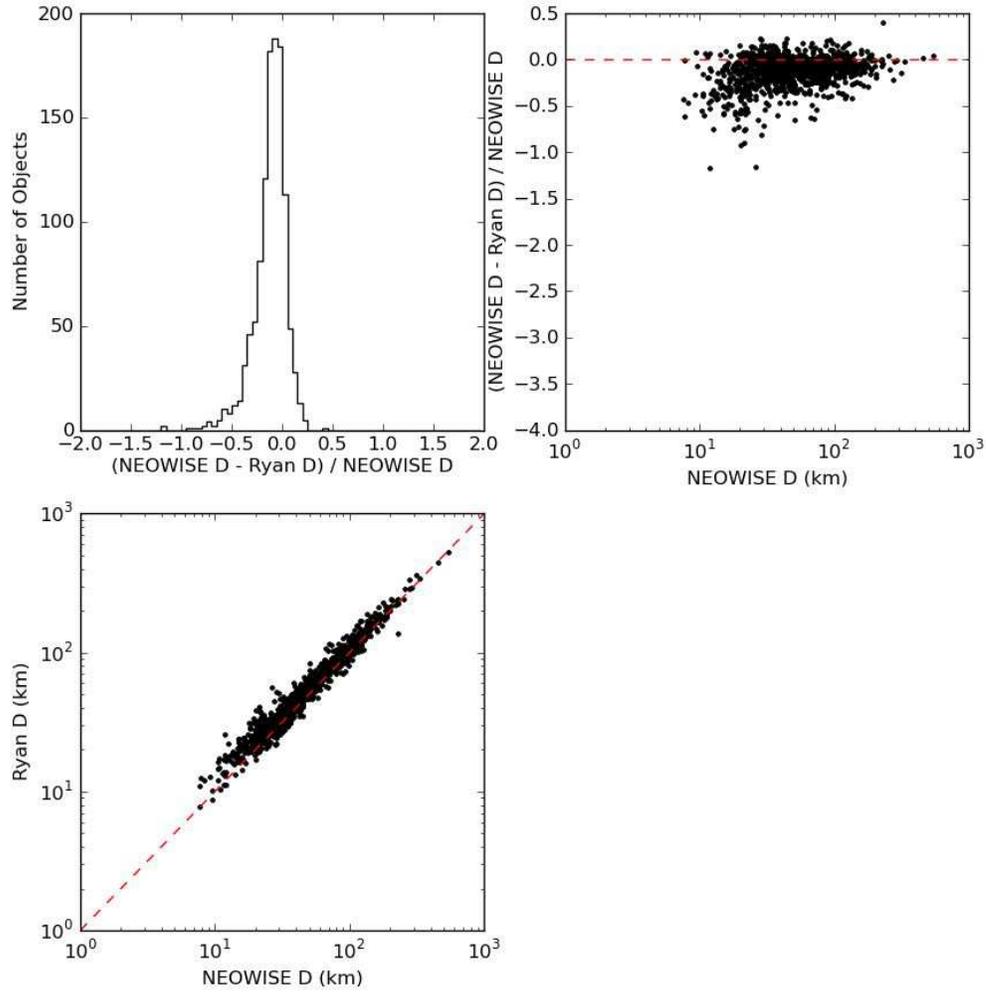}
\caption{\label{fig:ryan_diameter_diff}Comparison of the difference in diameters given by \citet{Ryan} and the diameters found by applying a spherical NEATM model to NEOWISE measurements for 1155 asteroids.  The diameters found by \citet{Ryan} tend to be larger than those found by NEOWISE.  
}
\end{figure}

\begin{figure}
\figurenum{3}
\includegraphics[width=6in]{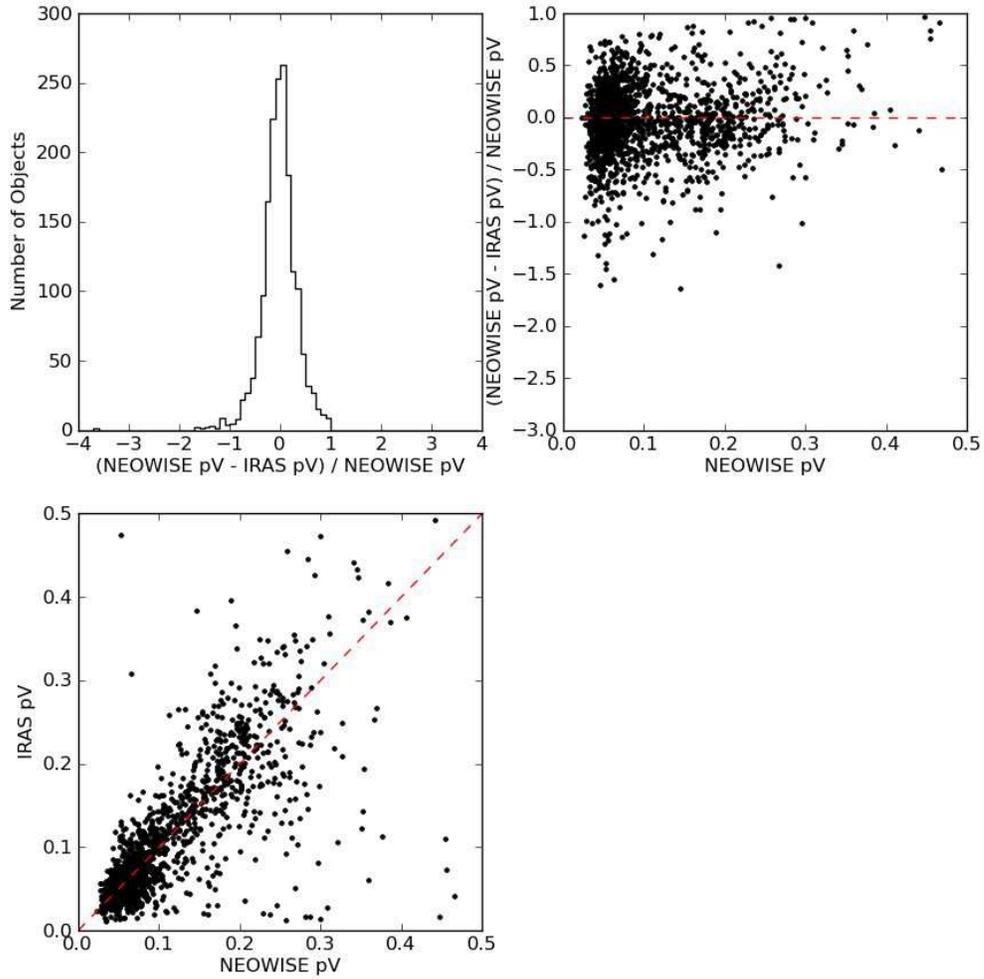}
\caption{\label{fig:iras_albedo_diff}Comparison of the difference in albedos given by \citet{TedescoPDS} and the diameters found by applying a spherical NEATM model to NEOWISE measurements for 1742 asteroids.  It can be seen that the values from \citet{TedescoPDS} tend to be slightly higher than NEOWISE-derived albedos.
}
\end{figure}

\begin{figure}
\figurenum{4}
\includegraphics[width=6in]{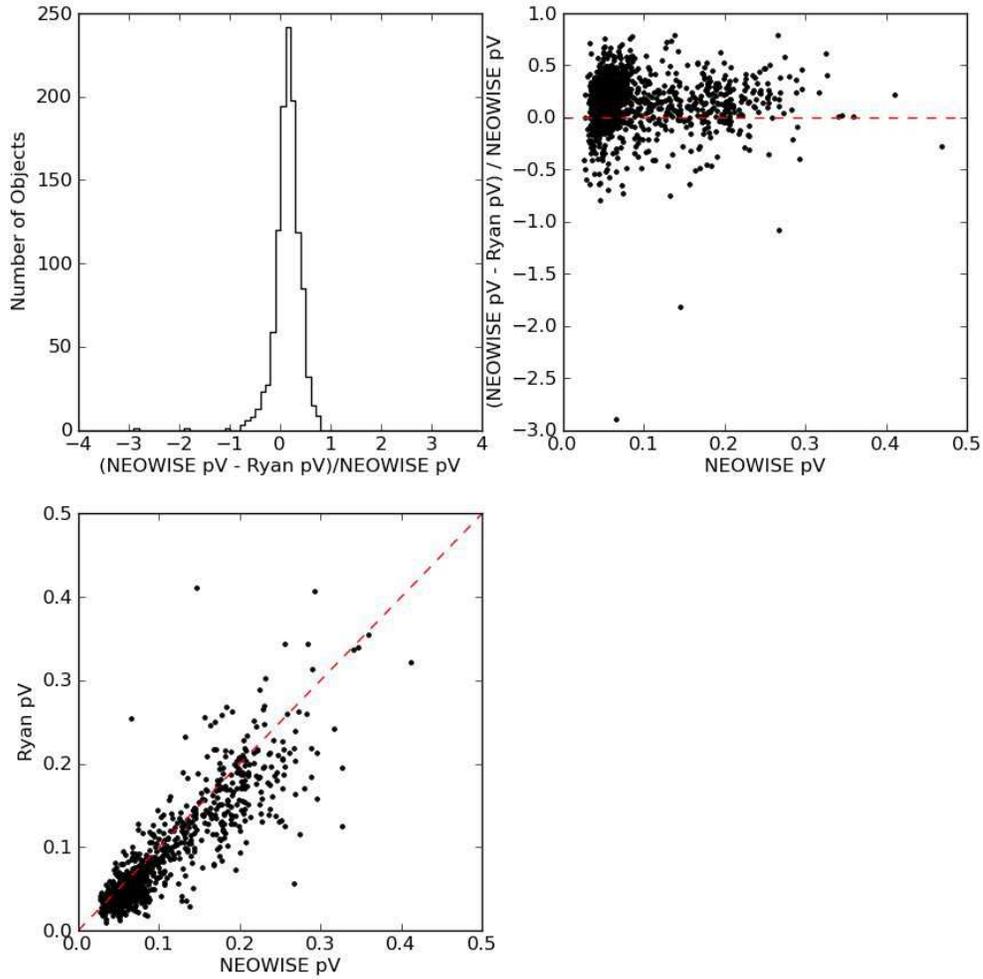}
\caption{\label{fig:ryan_albedo_diff}Comparison of the difference in albedos given by \citet{Ryan} and the albedos found by applying a spherical NEATM model to NEOWISE measurements for 1155 asteroids.  The albedos found by \citet{Ryan} tend to be lower than those found by NEOWISE. 
}
\end{figure}

\begin{figure}
\figurenum{5}
\includegraphics[width=6in]{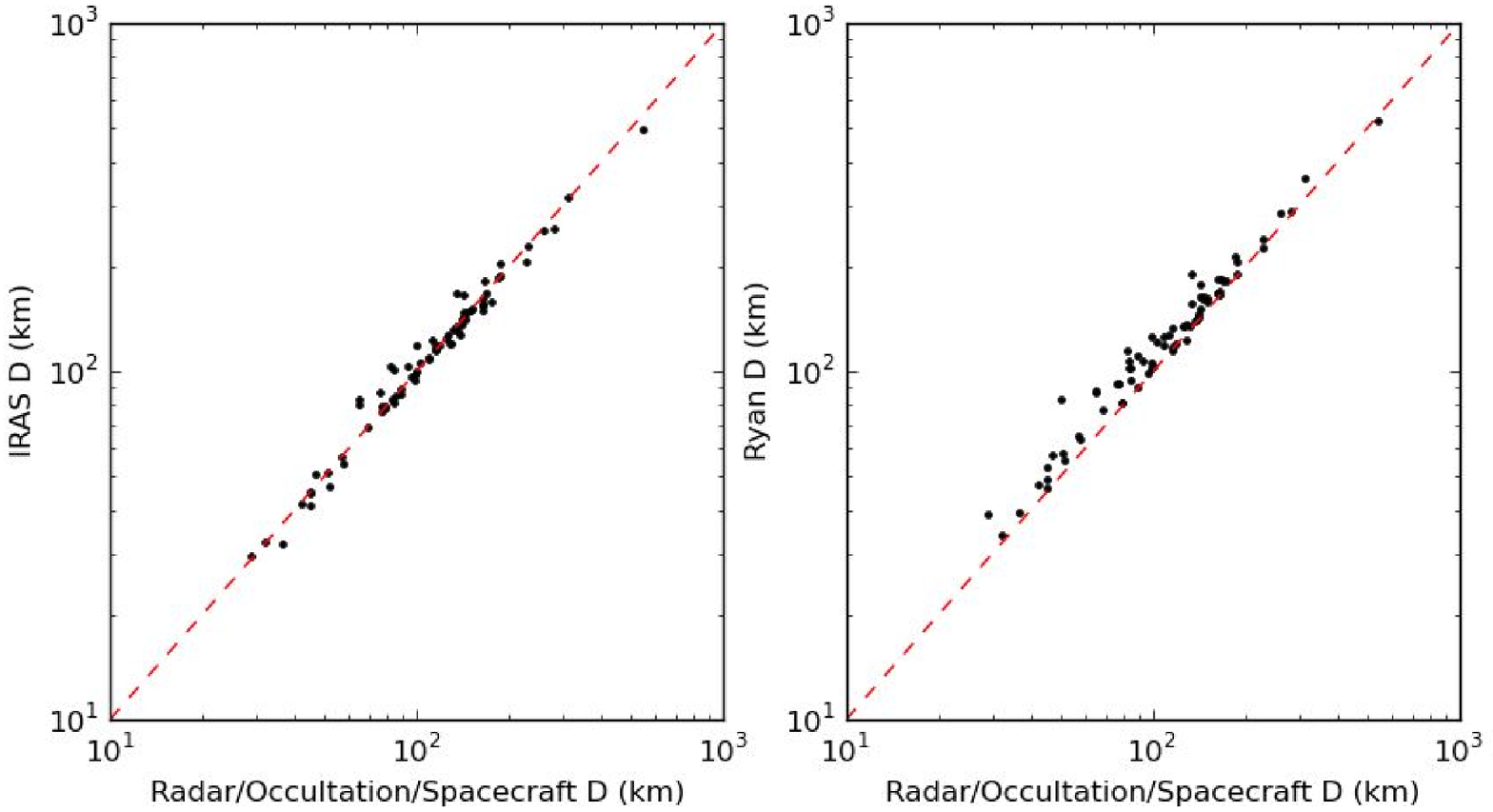}
\caption{\label{fig:radar}Comparison of diameters found by radar, stellar occultation, or in situ spacecraft imaging for 78 asteroids and the diameters found by \citet{TedescoPDS} and \citet{Ryan}.  The objects were selected to have $W3$ peak-to-peak amplitudes $<$0.5.  It can be seen that both methods of computing IRAS diameters are somewhat biased when compared to the radar diameters.  In \citet{Mainzer11b}, we show that there is no systematic bias found in model and observed WISE magnitudes for a similar set of objects with independently measured diameters.
}
\end{figure}

\citet{Tedesco02} reported observations of $\sim$2200 asteroid within the Supplemental IRAS Minor Planet Survey (SIMPS).  The diameters and albedos in that work were computed using the Standard Thermal Model.   Similar to the work of \citet{Walker}, who computed NEATM diameters and albedos for 654 asteroids using IRAS fluxes, \citet{Ryan} applied a spherical NEATM model to $\sim$1500 asteroids found in the SIMPS and Mid-Course Space Experiment \citep[MSX;][]{TedescoMSX} data.  Figures \ref{fig:iras_diameter_diff} and \ref{fig:ryan_diameter_diff} show the comparison between diameters found using NEOWISE observations and those given by \citet{TedescoPDS} and \citet{Ryan}, respectively.  Figures \ref{fig:iras_albedo_diff} and \ref{fig:ryan_albedo_diff} show the comparison between NEOWISE and Tedesco/Ryan albedos.  The diameters given by \citet{TedescoPDS} are systematically lower than the NEOWISE diameters, while the diameters computed by \citet{Ryan} are systematically higher.  Also, \pv\ given by \citet{TedescoPDS} is systematically higher than the NEOWISE \pv\ values, but the \citet{Ryan} values tend to be slightly lower.  In Figure \ref{fig:radar}, we show the diameters of \citet{TedescoPDS} and \citet{Ryan} compared to the diameters of 78 objects with diameters measured by radar, occultation, or spacecraft imaging (see \citet{Mainzer11b} for a complete list of references).  These objects have been selected to have $W3$ peak-to-peak amplitudes $<$0.5 magnitudes to avoid the worst complications caused by applying spherical models to highly elongated objects.  As can be seen in the figure, the \citet{TedescoPDS} diameters are skewed toward slightly smaller sizes than the radar, etc. measurements, whereas the \citet{Ryan} diameters are biased toward larger sizes.  We propose that these biases are caused by the band-to-band color corrections derived for IRAS.  As described in \citet{Tedesco02}, band-to-band corrections were derived by requiring that the 10 and 20 $\mu$m IRAS observations simultaneously matched a diameter derived from a stellar occultation of (1) Ceres.  Because the STM parameters were set to match Ceres, \citet{Tedesco02} found a 7\% difference between diameters derived for 13 objects with diameters measured from stellar occultations.  The NEATM model applied by \citet{Ryan} does not appear to address this issue.  Furthermore, \citet{Ryan} applied cutoffs to $\eta$ of 0.75 and 2.75.  As described above, in our NEATM models applied to the NEOWISE data, we cut off $\eta$ at the limit predicted by the Fast Rotating Model, $\pi$.  In the STM, $\eta$ was set to 0.756 by matching the diameters of Ceres and Pallas; however, there is no reason in principle why it cannot range to lower values.  In our application of NEATM, we allow $\eta$ to range arbitrarily low and find no fitted $\eta$ values less than $\sim$0.53. 

As described above, in \citet{Mainzer11b}, we used asteroids with previously known diameters and $H$ magnitudes to compare our model and observed WISE magnitudes, and we found no systematic offsets.  We conclude that the diameters for minor planets derived from NEOWISE are in generally good agreement with those found by IRAS and are likely more free of systematic biases than the diameters provided in either \citet{Tedesco02} or \citet{Ryan}.  Together with \cite{Mainzer11b}, this  demonstrates that the NEOWISE dataset will produce good quality physical parameters for the $>$157,000 minor planets it contains.

\section{Acknowledgments}

\acknowledgments{This publication makes use of data products from the \emph{Wide-field Infrared Survey Explorer}, which is a joint project of the University of California, Los Angeles, and the Jet Propulsion Laboratory/California Institute of Technology, funded by the National Aeronautics and Space Administration.  This publication also makes use of data products from NEOWISE, which is a project of the Jet Propulsion Laboratory/California Institute of Technology, funded by the Planetary Science Division of the National Aeronautics and Space Administration. We gratefully acknowledge the extraordinary services specific to NEOWISE contributed by the International Astronomical Union's Minor Planet Center, operated by the Harvard-Smithsonian Center for Astrophysics, and the Central Bureau for Astronomical Telegrams, operated by Harvard University.  We acknowledge use of NASA's Planetary Data System.  This research has made use of the NASA/IPAC Infrared Science Archive, which is operated by the Jet Propulsion Laboratory, California Institute of Technology, under contract with the National Aeronautics and Space Administration. This research has made use of NASA's Astrophysics Data System.}

\clearpage

\end{document}